\documentclass[sigconf]{acmart}

\AtBeginDocument{%
  \providecommand\BibTeX{{%
    \normalfont B\kern-0.5em{\scshape i\kern-0.25em b}\kern-0.8em\TeX}}}

\usepackage{multirow}
\usepackage{kotex}
\usepackage[normalem]{ulem}
\begin{document}

\title{MC$^2$SleepNet: Multi-modal Cross-masking \\ with Contrastive Learning for Sleep Stage Classification}


\author{Younghoon Na}
\email{yh0728@snu.ac.kr}
\affiliation{
  \institution{Seoul National University}
  \city{Seoul}
  \country{South Korea}
}

\author{Hyun Keun Ahn}
\email{hahn1123@gmail.com}
\affiliation{
  \institution{Seoul National University}
  \city{Seoul}
  \country{South Korea}
}

\author{Hyun-Kyung Lee}
\email{hyunkyung913@snu.ac.kr}
\affiliation{
  \institution{Seoul National University}
  \city{Seoul}
  \country{South Korea}
}

\author{Yoongeol Lee}
\email{crabyg71@gmail.com}
\affiliation{
  \institution{Seoul National University}
  \city{Seoul}
  \country{South Korea}
}

\author{Seung Hun Oh}
\email{gnsgus190@gmail.com}
\affiliation{
  \institution{Hallym University}
  \city{Chuncheon}
  \country{South Korea}
}

\author{Hongkwon Kim}
\email{kylekim00@gmail.com}
\affiliation{
  \institution{Hallym University}
  \city{Chuncheon}
  \country{South Korea}
}

\author{Jeong-Gun Lee}
\email{jeonggun.lee@hallym.ac.kr}
\affiliation{
  \institution{Hallym University}
  \city{Chuncheon}
  \country{South Korea}
}



\begin{abstract}

Sleep profoundly affects our health, and sleep deficiency or disorders can cause physical and mental problems. 
Despite significant findings from previous studies, challenges persist in optimizing deep learning models, especially in multi-modal learning for high-accuracy sleep stage classification. 
Our research introduces MC$^2$SleepNet (Multi-modal Cross-masking with Contrastive learning for Sleep stage classification Network). It aims to facilitate the effective collaboration between Convolutional Neural Networks (CNNs) and Transformer architectures for multi-modal training
with the help of contrastive learning and cross-masking.
Raw single channel EEG signals and corresponding spectrogram data provide differently characterized modalities for multi-modal learning.
Our MC$^2$SleepNet has achieved \textit{state-of-the-art} performance with an accuracy of both 84.6\% on the SleepEDF-78 and 88.6\% accuracy on the Sleep Heart Health Study (SHHS).
These results demonstrate the effective generalization of our proposed network across both small and large datasets.
%

\end{abstract}






\maketitle

\section{Introduction}

Sleep is crucial for human life, as it influences physical and mental health. Sleep disorders and deficiencies can profoundly affect an individual's overall well-being \cite{wulff2010sleep}.
Automating sleep scoring commonly relies on Polysomnography (PSG), a widely implemented key diagnostic tool, enabling the analysis of sleep disorders through the recording of physiological signals such as the electroencephalogram (EEG) for brain activity, Electrooculogram (EOG) for eye movement, and Electromyogram (EMG) for muscle activity, etc. 
These signals are broken down into 30-second intervals, known as epochs, to categorize sleep stages.

Every epoch is systematically classified using criteria from the Rechtschaffen and Kales (R\&K)~\cite{rechtschaffen1968manual} manual, and the more recent American Academy of Sleep Medicine (AASM)~\cite{berry2012aasm} standards. Both offer detailed guidelines for sleep stage classification.


Despite the success of PSG-based manual sleep diagnostics, due to extensive recording time, a labor-intensive labeling process, and the inherent inconsistency in labeling \cite{chapotot2010automated}, automating sleep scoring becomes essential to address these challenges effectively. \cite{10.5665/sleep.2548} work highlights potential problems and demonstrates that automatic sleep stage classification achieves performance comparable to human experts, requiring only a few seconds for labeling. 


\begin{figure*}[htb]
\centerline{\includegraphics[width = \textwidth]{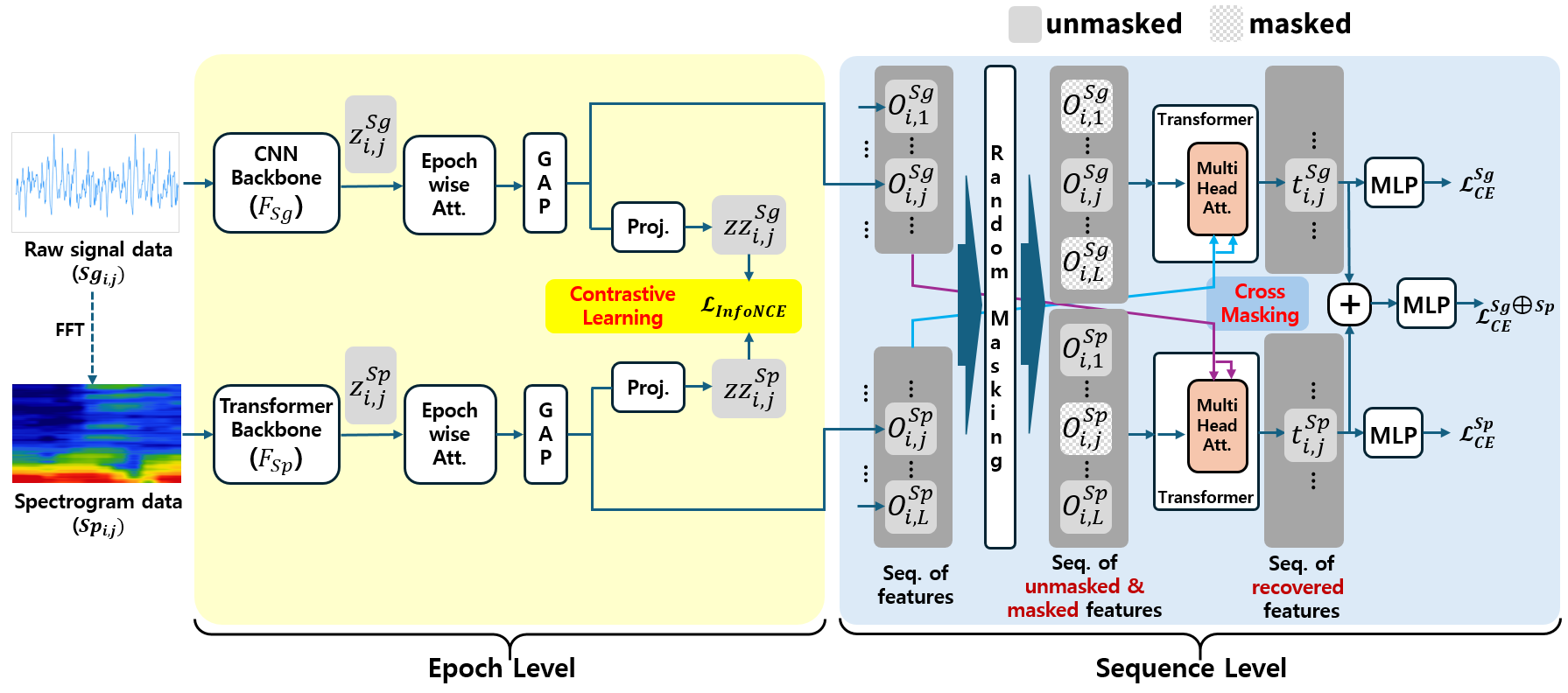}
}

\caption{ 
The MC$^2$SleepNet processes both raw signals and spectrograms as input. The raw signals are passed through the CNN-based backbone, while the spectrograms are fed into a Transformer-based backbone.
We carry out the pre-training steps concurrently across the granularity of epochs and sequences. To mitigate potential discrepancies between the features obtained from the data of each modality, our MC$^2$SleepNet employs InfoNCE loss.
Then, a random masking strategy with 50\% probability forces the model to refer to other features from other modality data through the cross-attention layers.
} 

\label{fig1}
\vspace{-5pt}
\end{figure*}


Recently, a series of studies, including DeepSleepNet \cite{supratak2017deepsleepnet}, SeqSleepNet \cite{phan2019seqsleepnet}, XSleepNet \cite{phan2021xsleepnet} introduced deep learning models that demonstrate state-of-the-art single channel performance in PSG datasets.
DeepSleepNet employs dual CNN models that capture temporal and frequency domain features with raw signal samples. 
SeqSleepNet introduces an RNN-based sequence-to-sequence network for processing spectrogram data.
In contrast to the DeepSleepNet architecture, to implement multi-modality training (raw signal and the corresponding spectrogram data), XSleepNet adopts hierarchical CNNs as a backbone network for raw signals and SeqSleepNet as a backbone network for spectrogram.
XSleepNet simultaneously optimizes two sub-models, leveraging a technique known as gradient blending \cite{what_makes_training_muli-modal_hard} for updating their gradients individually. 
Even though XSleepNet pioneers tried to optimize multi-modal tasks for sleep stage classification,
%
its effectiveness is constrained to specific deep-learning environments.


Due to their ease of optimization, there are various single-modal
%
training methods whose performance is on par with that of human experts. 
Within the context of multi-modal training, only XSleepNet holds a competitive status, still retaining its reliance on old-fashioned RNN-based architecture. This underscores inherent challenges in optimizing multi-modal tasks.
Recently, START \cite{10385393} and CoReSleepNet \cite{kontras2023coresleep} have incorporated multi-channel inputs using distinct networks, highlighting the potential for integrating diverse information and reducing the modality gap.
However, their contributions only display combined homogeneous information like same shape vectors consisting of different channels that use different models even though they show prominent results. They have yet to fully address the challenge of effectively combining heterogeneous information like signals and spectrograms.

In our approach, instead of utilizing gradient blending and old-fashioned RNN-based architectures, we develop a multi-modal model
combining the features extracted from multi-modal data.
%
Our network is designed with dual backbones, for the dual modalities of data samples: (1) ``Raw Signal View" and (2) ``Spectrogram View". 
A CNN-based backbone and a Transformer-based backbone are used as feature extractors for raw signal data and spectrograms, respectively. 
%

The proposed model architecture, {\bf M}ulti-modal {\bf C}ross-masking with {\bf C}ontrastive learning for {\bf Sleep} stage classification {\bf Net}work (MC$^2$SleepNet), is presented in Fig. \ref{fig1}. 
Training the MC$^2$SleepNet model is conducted in two steps: (1) Pre-training both ``{\bf Epoch- Level}" and  ``{\bf Sequence-Level}'' simultaneously with the use of both the supervised and self-supervised learning and (2) Fine-tuning the model while freezing the CNN and Transformer backbones trained during the pre-training step. 


During the pre-training step at an epoch-level, {\bf InfoNCE} loss \cite{oord2019representation} is employed to align and fine-tune the embedding from different network architectures, potentially improving the performance on downstream tasks. 
In the sequence-level, We introduce a novel concept, ``Cross-Masking", for implementing cross-attention to unmasked and masked pairs referring to \cite{NIPS2014_ImprovedMM} work.
It is noteworthy that two self-supervised learning techniques, ``Contrastive Learning" and "Masking Prediction", are integrated within a pre-training process. 
This integration encourages two different CNN and Transformer backbones to cross-examine the features extracted from another modality. It enables the inference of information that cannot be perceived by one view alone. 

In the fine-tuning stage, we freeze the backbone and update only the part of the model for sequence-level training. 
This process effectively leverages precise estimation by training inferring capability of the ground information from masked information \cite{NIPS2013_bengio}, while maintaining the integrity of the core representations.

Our proposed model shows state-of-the-art accuracy on the SleepEDF-78 (84.6\%) and SHHS (88.6\%) datasets, demonstrating improved classification performance compared to recently proposed deep learning models.

\section{Preliminary and Related Work}


\subsection{Spectrogram}
\label{Preliminary}

\begin{figure}
\centerline{\includegraphics[width = 230pt]{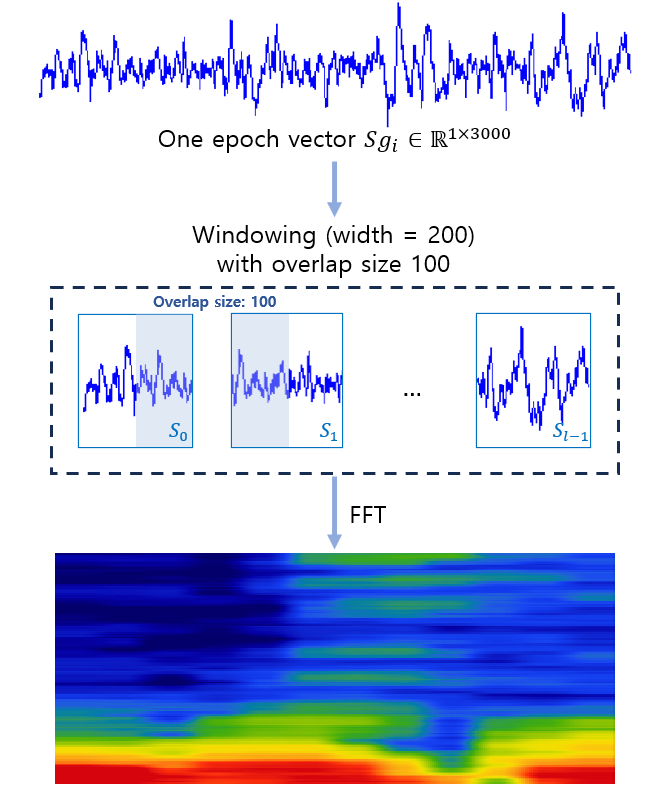}} 
\caption{A process for generating a spectrogram from a raw signal.}

\label{how_to_make_spectrogram}
\vspace{-15pt}
\end{figure}

A spectrogram is a two-dimensional representation that describes how the frequency spectrum of a signal changes over time.
Recently, spectrograms have been widely adopted across numerous application fields, including sleep stage classification.
%
%
Fig. \ref{how_to_make_spectrogram} shows how we obtain a sample of spectrogram data from an epoch of raw EEG signal data. We use single-channel EEG signals, ``Fpz-Cz" in a SleepEDF-78 dataset and ``C4-A1" in a SHHS dataset.
In this work, the employed signal is denoted by $\text{Sg}_{i,j} \in \mathbb{R}^{C \times 3000}$, where $C=1$. The signals are sampled at 100Hz and segmented into standard epoch samples of 30 seconds (100 data points $\times$ 30 sec = 3000 data points).
This approach aligns with datasets like SHHS, where signals originally sampled at 125Hz are resampled to 100Hz.

Then, we apply Fast Fourier Transforms (FFT) to the signals while adopting a window size of 200 data points (double the frequency) and an overlap of the equal size (100 data points), as recommended in studies like \cite{phan2021xsleepnet,kontras2023coresleep} as shown in Fig. \ref{how_to_make_spectrogram}. So, we have \text{29} time segments for an epoch, and each segment is transformed into a frequency spectrum, which is then normalized between 0 and 128. This setup efficiently captures the spectral characteristics of the EEG data. 

As a post-FFT processing, we implement log-magnitude transformations to normalize variance. Then we have a spectrogram data sample, $\text{Sp}_{i,j} \in \mathbb{R}^{C \times T \times \text{Freq}}$ where $T$ is the number of time segments (it is set to 29) and $\text{Freq}$ is set to 129.

\subsection{Sleep stage classification} 

Sleep stage classification can be leveraged for diagnosing sleep deficiency or disorders. In this classification, 30-second epochs are classified into several classes. According to the AASM guidelines \cite{berry2012aasm}, there are five classes for the sleep stages: Wake, NREM1, NREM2, NREM3, and REM. The datasets used in this work, SleepEDF-78 and SHHS datasets are labeled following the AASM guidelines.

%


As an initial deep learning model for sleep stage classification, CNN-based works are used to recognize patterns in biological signals \cite{sors2018convolutional,phan2018joint} which effectively capture features of each epoch data.
On the other hand, Recurrent Neural Networks (RNN) are used to extract sequential features from continuous sleep cycles and predict transitions between stages \cite{michielli2019cascaded,mousavi2019sleepeegnet}. To better exploit the sequential features within a consecutive sequence of epochs, variants such as Long-Short-Term Memory (LSTM) \cite{hochreiter1997long} and Gated Recurrent Units (GRU) \cite{chung2014empirical} are particularly prevalent in sleep stage classification.
Those methods, which employ a one-to-one approach \cite{Phan2018_DNNwith1MaxPooling}, suffer from primary drawbacks such as increased training time.

To enhance the efficiency of training steps, a many-to-many framework, such as SleepNet \cite{biswal2017sleepnet}, has been leveraged to predict the loss of multiple sequences.
DeepSleepNet \cite{supratak2017deepsleepnet} combines the feature extraction capability of CNN with the sequential data interpretation power of BiLSTMs. While it excels in deciphering complex datasets, it has trade-offs such as a large footprint and extended training duration.
However, the model demands intricate hyperparameter adjustments and incurs high computational expenses.

In \cite{oropesa1999sleep}, spectrogram was first adopted into various methods. Among them, SeqSleepNet \cite{phan2019seqsleepnet} effectively navigated long-term dependencies within spectrograms and attention mechanisms.
SleepTransformer \cite{phan2022sleeptransformer} used self-attention to understand the temporal complexities of spectrograms. Its ability to process in parallel speeded up training, but the necessity for large datasets and challenges in model interpretability were its notable drawbacks. 

For example, SleepTransformer had shown prominent results in large-size datasets such as SHHS, but relatively poor performance in small-size datasets such as SleepEDF. 
Consequently, numerous researchers have opted for RNN-based networks for spectrograms instead of Transformer-based models, as seen in works such as L-SeqSleepNet \cite{phan2023seqsleepnet}, XSleepNet, MVFSleepNet \cite{li2022mvf}. Although attempts have been made to utilize Transformer-based architectures, such as in CoReSleepNet \cite{kontras2023coresleep}, they have primarily been applied to large-scale datasets.

To date, no research has surpassed the XSleepNet architecture in multi-modal learning in sleep stage classification.
%
We expect that incorporating multi-modal data, including spectrograms, into a model combined with a Transformer and self-supervised learning will robustly enhance performance. This approach could potentially address difficult optimization problems more effectively.


\subsection{Self-supervised learning in sleep stage classification} 

Self-supervised learning was introduced with a specific focus on defining positive and negative samples \cite{9533305} and contrastive loss \cite{DBLP:journals/corr/abs-1807-03748} is used for a single-channel EEG. Particularly, CoSleep utilized InfoNCE loss in the multi-view learning era \cite{CoSleep_IEEE2021_multiview_with_CL}. Both models demonstrated enhanced performance and also showed their competitiveness on smaller datasets.
On the other hand, MAEEG \cite{chien2022maeeg} and MaskSleepNet \cite{MaskSleepNet2023_IEEE} adopted a masking strategy to extract information in the frequency domain while being trained to effectively reconstruct the original raw signal.
These whole networks were trained in a pre-training step to maximize their performance.

In addition to the epoch-level training, exploiting sequential features embedded within epoch sequences is crucial for achieving even higher classification accuracy. 
To train the sequential features from a sequence of multiple epoch samples, BENDR \cite{kostas2021bendr} uses a Transformer and contrastive self-supervised learning. 

From the work proving that a masking strategy can enhance the performances \cite{NIPS2013_bengio}, we expect that the use of the masking strategy for sequence-level training can lead to performance improvement.
However, BENDR demonstrates that performance improvement in self-supervised learning can only be achieved when there is a large amount of data samples available to capture their relationships.

\section{Methodology}\label{Methodology}

Given a training dataset $DS = \{SS_i\}_{i=1}^{N}$ has $N$ samples and each sample, $SS_i$ is an ordered tuple of sequential $L$ epochs.
Then $SS_i$ can be denoted by $(S_{i, j})_{j=1}^{L}$, in which $S_{i,j}$ corresponds to the $j$-th epoch of $SS_i$ and $L$ is set to 21 in this work.
The $S_{i,j}$ is a 3-element tuple and it is denoted by $(\text{Sg}_{i,j}, \text{Sp}_{i,j}, {y}_{i,j})$.
%
The first two elements of a tuple correspond to multi-modality data of an epoch: a raw signal data $(\text{Sg}_{i,j})$ and a spectrogram data $(\text{Sp}_{i,j})$. The last element, ${y}_{i,j}$, is a ground-truth label for the epoch.
Specifically, each sample of the training datasets consists of
$SS_i = ((\text{Sg}_{i,1}, {\text{Sp}_{i,1}}, {y}_{i,1}), \cdots, (\text{Sg}_{i,L}, {\text{Sp}_{i,L}}, {y}_{i,L})). $
The label ${y}_{i,j}$ $\in$ $\{$Wake, NREM1, NREM2, NREM3, REM$\}$ represents a stage class for $S_{i,j}$, $\text{Sg}_{i,j} \in \mathbb{R}^{C \times 3000}$ represents 30-second information across $C$ channels sampled at 100Hz frequency. 
In this work, we consider a single channel EEG so that $C=1$.

The overall model architecture and its layer components are presented in Fig \ref{fig1}. We use a CNN backbone network denoted by $\textbf{F}_{Sg}$ to draw out features from the raw signal data. Another backbone, $\textbf{F}_{Sp}$, for the spectrogram data, is based on a Transformer network.


\subsection{Epoch level training}\label{3-1_epoch_level_training} 

In the backbone for raw signal data, following the \cite{perslev2019utime,SEO2020102037,lee2024sleepyco}, we adopt 5 CNN blocks with the following channel configurations: Each of the first two blocks has two convolutional layers, while the subsequent blocks each have three convolutional layers. Maxpool layers with a width of `5' are appended after all the blocks except the first block.
In the whole CNN blocks, we adopt a kernel size of `3' and the same padding size with the stride of `1'. 
Finally, we implemented a maxpool layer with a width size of `2' to adjust the CNN output dimensions that match the Transformer output dimensions.
%

The backbone model function $\mathbf{F}_{Sg}$ produces the feature having the raw signal $\mathbf{Z}^{Sg}_{i, j}$ as its domain, making dimension of the feature to $\mathbf{F}_{Sg}(\text{Sg}_{i, j}) \in \mathbb{R}^{\text{batch} \times 5 \times 128}$.
On the other hand, as described in Section \ref{Preliminary}, a spectrogram data sample, $\text{Sp}_{i,j} \in \mathbb{R}^{C \times 129 \times 29}$, undergoes a simple CNN-based resizing layer and then the output of the resizing layer is further augmented with positional encoding vector (PE) to yield $\text{Sp}^*_{i,j} \in \mathbb{R}^{C \times 29 \times 128}$ as defined in Eq. \ref{eq1}.
This resizing aims to exploit the potential benefits of dimensionality being a power of two and provide additional non-linearity.

\vspace{-10pt}

\begin{align}
\label{eq1}
\text{Sp}^*_{i, j} = \text{proj}_{CNN}(\text{Sp}_{i, j}) + \text{PE}
\end{align}

In the backbone of the spectrogram, $\textbf{F}_{Sp}$, a Transformer layer performs a sequence of the computations defined from Eq. \ref{2. attn-1} to Eq. \ref{2. attn-7}.
Specifically, $\text{Attn}(x)$ in Eq. \ref{2. attn-1} uses a set of typical learnable parameters for a Transformer: $\mathbf{W}^K_h$, $\mathbf{W}^Q_h$, $\mathbf{W}^V_h$ $\in$ $\mathbb{R}^{d_{model} \times \frac{d_{mat}}{d_{head}}}$, $\mathbf{W}^{F}_{1}$ $\in$ $\mathbb{R}^{d_{model} \times d_{FF}}$, $\mathbf{W}^F_{2} \in \mathbb{R}^{d_{FF} \times d_{model}}$, $\mathbf{b_1} \in \mathbb{R}^{d_{FF}}$, and $\mathbf{b_2} \in \mathbb{R}^{d_{model}}$. We set $d_k = 16$ and $d_{FF}$ is set to 1024 with a dropout rate of 0.1. 

$\text{Sp}^{*}_{i,j}$ is given as an input, $\mathbf{Z}^{1}_{i, j}$, for the first Transformer layer as presented in Eq. \ref{2. attn-7} with $l=1$.
After `$iter$' (4 is used as $iter$) iterations of the Transformer layer, the output, $\mathbf{Z}^{iter+1}_{i, j}$, is produced as a shape of $\mathbb{R}^{\text{batch} \times 5 \times 128}$ with $d_{head} = 8$. The final output of Eq. \ref{2. attn-7} can be re-written as $\mathbf{Z}^{Sp}_{i, j}$.


\vspace{-15pt}
\begin{align}
\label{2. attn-1}
\text{Attn}(x) = \sigma\left(\frac{\mathbf{W}^K_h x \cdot (\mathbf{W}^Q_h x)^ \top}{\sqrt{d_k}} \mathbf{W}^V_h x\right) 
\end{align}
\vspace{-20pt}

\begin{align}
\label{2. attn-2}
\text{SA}(x) = \text{concat}[\text{Attn}_1(\ldots), \ldots, \text{Attn}_H(x)]
\end{align}
\vspace{-29pt}

\begin{align}
\label{2. attn-3}
\text{FF}(x) = \max(0, x \mathbf{W}^F_1 + \mathbf{b}^F_1) \mathbf{W}^F_2 + \mathbf{b}^F_2 
\end{align}
\vspace{-29pt}
 
\begin{align}
\label{2. attn-4}
\text{LN}(\text{F}(x)) = \text{layernorm}(x + \text{F}(x))
\end{align}
\vspace{-29pt}

\begin{align}
\label{2. attn-5}
\text{SA Block}(x) = \text{LN}(\text{SA}(x))
\end{align}
\vspace{-29pt}

\begin{align}
\label{2. attn-6}
\text{TF}(x) = \text{SA Block}(x) + \text{LN}(\text{FF}(\text{SA Block}(x)))
\end{align}
\vspace{-29pt}

\begin{align}
\label{2. attn-7}
\text{for } l = 1 \text{ to } iter: \mathbf{Z}^{1+1}_{i,j} = \text{TF}(\mathbf{Z}^{l}_{i,j}) 
\end{align}
\vspace{-10pt}


Both the raw signal and spectrogram inputs are fed into their corresponding backbone networks independently as presented in Fig. \ref{fig1}. After getting $\textbf{Z}^{Sp}_{i, j}$, an epoch-wise attention ($\textbf{EW}$) defined in Eq. \ref{3. EW-attn-3} is applied to enhance features further.  
Attention, $\alpha_t$, for {\bf EW} is defined by Eq. \ref{3. EW-attn-1} and Eq. \ref{3. EW-attn-2}.
In Eq. \ref{3. EW-attn-1}, $\mathbf{W} \in \mathbb{R}^{A \times F}$ and $\mathbf{b}_a \in \mathbb{R}^{A}$ are learnable parameters and $A$ is an attention size. The parameters, $A$ and $F$, are set to 128.

The epoch-wise attention operation is also applied to the feature, $\textbf{Z}^{Sg}_{i, j}$, extracted from raw signal data to accentuate the important channel information in the feature.
In Eq. \ref{3. EW-attn-1}, $\mathbf{z}_t$ can be $\textbf{Z}^{Sg}_{i, j}$ or $\textbf{Z}^{Sp}_{i,j}$ according to the input. The index, $t$, of $z_t$ has it range, $1 \le t \le T$ where $T$ is `5' for $\textbf{Z}^{Sg}_{i, j}$ and `29' for $\textbf{Z}^{Sp}_{i, j}$.

\vspace{-10pt}
\begin{align}
\label{3. EW-attn-1}
\mathbf{a}_t = \tanh(\mathbf{W} \cdot \mathbf{z}_t + \mathbf{b}_a) 
\end{align}
\vspace{-27pt}

\begin{align}
\label{3. EW-attn-2}
\alpha_t = \frac{\exp(\mathbf{a}_t^ \top \mathbf{a}_e)}{\sum_{t=1}^{T} \exp(\mathbf{a}_t^ \top \mathbf{a}_e)}
\end{align}

Then, we employ global average pooling (GAP) to effectively condense the high-dimensional feature maps. This application of GAP results in compact representations of the feature vectors and finally we have $\textbf{O}^{Sg}_{i, j} \text{ and } \textbf{O}^{Sp}_{i, j} \in \mathbb{R}^{\text{batch} \times d_{model}}$ as depicted in Eq. \ref{3. EW-attn-4}.

\vspace{-15pt}
\begin{align}
\label{3. EW-attn-3}
\textbf{EW}(\mathbf{Z}_{i, j}) = \sum_{t=1}^{T} \alpha_{t} \mathbf{z}_t 
\end{align}
\vspace{-15pt}

\begin{align}
\label{3. EW-attn-4}
\textbf{O}^{Sg}_{i, j} &= \text{GAP}(\textbf{EW}(\textbf{Z}^{Sg}_{i, j})) \nonumber \\ 
\textbf{O}^{Sp}_{i, j} &= \text{GAP}(\textbf{EW}(\textbf{Z}^{Sp}_{i, j}))
\end{align}

After passing through the global average pooling layer,  InfoNCE loss is calculated for training while inducing an embedding feature space of similar semantic meanings. Before directly applying the InfoNCE loss, a projection layer is added to help a model be optimized by decoupling the layers before and after the projection layer.
The projection layer takes $\mathbf{O}^{Sg}_{i, j}$ and $\mathbf{O}^{Sp}_{i, j}$, and it generates $\mathbf{zz}^{Sg}_{i,j}$ and $\mathbf{zz}^{Sp}_{i,j}$, respectively as presented in Eq. \ref{eq12} 
($\mathbf{zz}^{Sg}_{i,j}$ and $\mathbf{zz}^{Sp}_{i,j}$ $\in \mathbb{R}^{d_{model} \times d_{proj}}$, $d_{proj} = 128$).
%

The loss terms for the InfoNCE are described in Eq. \ref{eq13}, Eq. \ref{eq14}, and Eq. \ref{InfoNCE_all}. We employ these loss equations to assess similarities between the features by contrasting the raw signal features with those of the spectrogram.
For effective contrast, negative samples are collected from a batch with a batch size of $B$.
Additionally, $\mathcal{L}_{sig2spc}$ and $\mathcal{L}_{spc2sig}$ are evaluated as the averages of the similarity ratios over the $L$ epochs of a sequence sample. Finally, our epoch-level loss function can be defined in Eq. \ref{InfoNCE_all}.

\begin{align}
\label{eq12}
\mathbf{zz}^{Sg}_{i,j} &= \text{Proj}(\textbf{O}^{Sg}_{i, j}) \quad  \mathbf{zz}^{Sp}_{i,j} = \text{Proj}(\textbf{O}^{Sp}_{i, j})
\end{align}
\vspace{-15pt}

\begin{align}
\label{eq13}
\mathcal{L}_{sig2spc} &= -\log \frac{1}{L}\sum_{j=1}^{L}\frac{\exp((\mathbf{zz}^{Sg}_{i,j})^ \top \cdot \mathbf{zz}^{Sp}_{i,j} / \tau)}{\sum_{k=1}^{B} \exp((\mathbf{zz}^{Sg}_{i,j})^ \top \cdot \mathbf{zz}^{Sp}_{k, j} / \tau)} \\
\label{eq14} \mathcal{L}_{spc2sig} &= -\log \frac{1}{L} \sum_{j=1}^{L}\frac{\exp((\mathbf{zz}^{Sp}_{i,j})^ \top \cdot \mathbf{zz}^{Sg}_{i,j} / \tau)}{\sum_{k=1}^{B} \exp((\mathbf{zz}^{Sp}_{i,j})^ \top \cdot \mathbf{zz}^{Sg}_{k, j} / \tau)}
\end{align}


\vspace{-10pt}
\begin{align}
\label{InfoNCE_all}
\mathcal{L}_{epoch-level} &= (\mathcal{L}_{sig2spc} + \mathcal{L}_{spc2sig})/2 \;\;\; 
\end{align}

Also, to mitigate unstable optimization and encourage learning useful features during the initial training steps, we employ pre-trained backbone networks for epoch-level training. 


\subsection{Sequence level training}\label{3-2_sequence_level_training} 

In our sequence-level training, we use a sequence of the features for each modality: 
$(\textbf{O}^{Sg}_{i, j})_{j=1}^L$ and $(\textbf{O}^{Sp}_{i, j})_{j=1}^L$ obtained from $SS_i$ with the epoch-level backbones, epoch-wise attention and GAP.
The masking strategy is integrated into the sequence-level training process to leverage the complementary nature of multi-modal characteristic, 
enabling features from one modality
to aid in restoring masked features of the other modality.
%
For random masking, we use a masking probability of 50\%. By selectively hiding half of the input, the training encourages a more robust learning process, intending to recover the masked features.

Our proposed method, named ``Cross-Masking'', allows our model to extract the sequential features from the neighboring tokens for each modality 
while extensively referencing the feature information from the different modalities 
through cross-attention layers.
For the Cross-Masking, two Transformer blocks are used for each modality 
and the Transformers exchange the feature information using a cross-attention mechanism.
Each Transformer block is composed of four Transformer layers. In the Transformer block for the Cross-Masking, the primary distinction from the Transformer backbone given in Eq. \ref{2. attn-7} is that a pair of cross-attention layer and a layer normalization is added after a self-attention layer and a layer normalization. 
The cross-attention in the Transformer block for raw signal features can be described in Eq. \ref{eq16} where $\textbf{O}^{Sg}_{i, j}$ is used as a query while $\textbf{O}^{Sp}_{i, j}$ is used as key and a value. Note that the order of arguments in the function of Eq. \ref{eq16} is important.






\newcommand{\vect}[1]{\mathbf{#1}}
\newcommand{\sg}{\vect{O}^{Sg}_{i, j}}
\newcommand{\spsp}{\vect{O}^{Sp}_{i, j}}
\newcommand{\whk}{\mathbf{W}^K_h}
\newcommand{\whq}{\mathbf{W}^Q_h}
\newcommand{\whv}{\mathbf{W}^V_h}
\newcommand{\dk}{\sqrt{d_k}}

\begin{align}
\text{Attn}(\sg, \spsp) &= \sigma\left(\frac{(\whk \sg) \cdot (\whq \spsp)^\top}{\dk} \whv \sg\right) \label{eq16}
\end{align}

For the cross-attention for the spectrogram features, the roles of $\textbf{O}^{Sg}_{i, j}$ and $\textbf{O}^{Sp}_{i, j}$ are interchanged so that $\textbf{O}^{Sp}_{i, j}$ is used as a query while $\textbf{O}^{Sg}_{i, j}$ is used as a key and a value.

The critical hyperparameter for the ``Cross-Masking'' is the masking ratio within the sequence of the features. Our random masking strategy is expected to enhance the model generalization by training a recovery process from incomplete or unseen data distributions as used in MAEEG \cite{chien2022maeeg} and BERT \cite{devlin2019bert}.
Even though the higher masking ratio is applied in paired samples, the model could compel to reference the features from different modalities 
through cross-attention layers to predict the label information. This results in more accurate predictions and facilitates the harmonization of different modalities.
Our masked tokens are shared and learnable within an expanded batch.


To further improve our model, a fine-tuning step is conducted after obtaining the pre-trained model with the masking strategy. By freezing the epoch-level backbone models, we focus on refining the sequence-level predictions by training a sequence model part with the pairs of the unmasked feature sequences, fostering a more cohesive and effective dual encoder architecture. 
Note that the masking is not used for the fine-tuning step. 
This strategic approach can eliminate the need for redundant adjustments at the epoch-level, streamlining the model's optimization for peak performance. 

A loss function for the recovering process is presented in Eq. \ref{eq17}. In the equation, $y_{i, j}$ in a set, $Y$, are the true labels in the $j$-th epoch data in the $i$-th samples across a sequence length of $L$ and a total of $N$ samples.
%
Similarly, $\hat{y}_{i, j}$ in a set, $\hat{Y}$, indicates the predicted label. The ground true label, $y_{i, j}$, and the predicted label, $\hat{y}_{i, j}$, are paired with the indices, $i$ and $j$.

\vspace{-1pt}
\begin{align}
\label{eq17}
\mathcal{L}_{recover}(\hat{Y}, Y) = -\frac{1}{N} \sum_{i=1}^{N} \sum_{j=1}^{L} \log(\hat{y}_{i, j}) \cdot y_{i,j} \\ \nonumber
\text{where } y_{i, j} \in Y \text{and } \hat{y}_{i, j} \in \hat{Y}
\end{align}
\vspace{-10pt}


The loss function for the sequence-level training can be expressed in Eq. \ref{eq21} and it consists of three cross-entropy loss functions. The detailed model structure for the sequence-level training is presented in the right part of the proposed model in Fig. \ref{fig1}.
%
%
The recovered features by our model are represented by $t^{Sg}_{i,j}$ or $t^{Sp}_{i,j}$.
A cross-entropy loss function is used for evaluating the quality of the recovered features as shown in Eq. \ref{eq20-1} and Eq. \ref{eq20-2}. In addition, we apply a cross-entropy function to the features obtained by concatenating the $t^{Sg}_{i,j}$ and $t^{Sp}_{i,j}$. The concatenate operator is denoted by $\oplus$ in Eq. \ref{eq20-3}. 
Then, those features are fed forward through a multi-layer perceptron (MLP) to generate the class predictions. 
%

\vspace{-5pt}
\begin{align}
\label{eq21}
\mathcal{L}_{sequence-level} = w_{1} \cdot \mathcal{L}^{Sg}_{CE} + w_{2} \cdot \mathcal{L}^{Sp}_{CE} + w_{3} \cdot  \mathcal{L}^{Sg \oplus Sp}_{CE}
\end{align}

\vspace{-10pt}
\begin{align}
\label{eq20-1}
\mathcal{L}^{Sg}_{CE} &= \mathcal{L}_{recover}(\hat{Y}^{Sg}, Y) \text{, where } 
\\ \nonumber
\text{MLP}(t^{Sg}_{i, j}) \in \hat{Y}&^{Sg}, \,\forall\, t^{Sg}_{i, j} : i \in \{1, \ldots, N\}, j \in \{1, \ldots, L\}
\end{align}

\vspace{-15pt}
\begin{align}
\label{eq20-2}
\mathcal{L}^{Sp}_{CE} &= \mathcal{L}_{recover}(\hat{Y}^{Sp}, Y) \text{, where } 
\\ \nonumber
\text{MLP}(t^{Sp}_{i, j}) \in \hat{Y}&^{Sp}, \,\forall\, t^{Sp}_{i, j} : i \in \{1, \ldots, N\}, j \in \{1, \ldots, L\}
\end{align}

\vspace{-15pt}
\begin{align}
\label{eq20-3}
\mathcal{L}^{Sg \oplus Sp}_{CE} &= \mathcal{L}_{recover}(\hat{Y}^{SS}, Y) \text{, where } 
\\ \nonumber
\text{MLP}(t^{SS}_{i, j}) \in \hat{Y}^{SS}, \,\forall\, t^{SS}_{i, j}&= t^{Sg}_{i, j} \oplus t^{Sp}_{i, j} \nonumber : i \in \{1, \ldots, N\} \text{ and } j \in \{1, \ldots, L\}
\end{align}
\vspace{-5pt}

In Eq. \ref{eq21}, $w_{1}$ is set to 1 while $w_{2}$ and $w_{3}$ are set to 0.1 for a pre-training step.
For a fine-tuning step, the sequence loss, denoted $\mathcal{L}_{sequence-level}$, given in Eq. \ref{eq21} is used but $w_{1}$, $w_{2}$ and $w_{3}$ are set to 1.
For an inference, we use the predictions generated by the MLP using the concatenated features.
Finally, our total loss functions for the pre-training step can be expressed as the sum of an epoch-level loss and a sequence-level loss as given in Eq. \ref{total_pre_train_loss_function}.

\vspace{-10pt}
\begin{gather}
\label{total_pre_train_loss_function}
\mathcal{L}_{pretrain} = \mathcal{L}_{epoch-level} + \mathcal{L}_{sequence-level}
\end{gather}

\section{Experiments} 

%
%


\subsection{Datasets and training setup}  

\begin{table*}
\centering
\caption{\textbf{SleepEDF-78 and SHHS dataset characteristics}}
\vspace{-10pt}
\label{table1}
\begin{tabular}{ccc|cccccc}
\hline\hline
Dataset & No. of Subjects & EEG Channel & Wake & N1 & N2 & N3 & REM & Total \\
\hline
\multirow{2}{*}{SleepEDF} & \multirow{2}{*}{78} & \multirow{2}{*}{Fpz-Cz} & 65,249 & 20,512 & 67,772 & 12,898 & 20,582 & \multirow{2}{*}{192,233} \\
                            &                     &                         & (33.9\%) & (10.7\%) & (35.3\%) & (6.7\%) & (13.4\%) &  \\

\multirow{2}{*}{SHHS}      & \multirow{2}{*}{5,791} & \multirow{2}{*}{C4-A1} & 1,609,066 & 217,404 & 2,396,336 & 739,228 & 817,295 & \multirow{2}{*}{5,779,329} \\
                            &                       &                        & (27.8\%) & (3.8\%) & (41.5\%) & (12.8\%) & (14.1\%) &  \\

\multirow{2}{*}{SHHS}      & \multirow{2}{*}{5,463} & \multirow{2}{*}{C4-A1} & 1,497,385 & 206,008 & 2,221,265 & 735,702 & 796,846 & \multirow{2}{*}{5,457,206} \\
                            &                       &                        & (27.4\%) & (3.8\%) & (40.7\%) & (13.5\%) & (14.6\%) &  \\
\hline\hline

\end{tabular}
\end{table*}

\subsubsection{Datasets:} \label{section4_dataset}
Two publicly available datasets are employed to assess the performance of the proposed model: SHHS and SleepEDF-78.
The SHHS database contains massive PSG records from 5,793 subjects aged 39-90. 
We use the preprocessing guidelines of the previous study and we utilize data from 5,791 patients in the SHHS dataset (it is named `SHHS 5791'). Additionally, we excluded the recordings that do not include all five sleep stages while following the approach detailed in \cite{SORS2018107} and the dataset is named `SHHS 5463'. 
%
%
We validate the dataset under two conditions: one with 5,791 patients and another with 5,463 patients. Detailed information on the two datasets is presented in Table \ref{table1}.

We use the SHHS dataset to test if our model works well with a large-size dataset. The PSG records of 100 patients are excluded from the dataset for validation and then the remaining dataset is divided into training data and testing data with a partitioning ratio of 70:30. 
On the other hand, the SleepEDF-78 contains 78 healthy Caucasian subjects aged 25-101, and it has smaller PSG records than those of the SHHS.
We use the SleepEDF-78 as our small dataset to test if our model works well on a small dataset. We validate our model using the SleepEDF-78 through 10-fold cross-validation. 



\subsubsection{Training setup:} In all experiments, the networks were trained using an Adam optimizer \cite{kingma2017adam} with the learning rate ($lr$) of $5 \times 10^{-4}$, $\beta_1 = 0.9$, $\beta_2 = 0.999$ and the weight decay of $1 \times 10^{-5}$ to prevent over-fitting problems. We adopt $B=32$ as mini-batch sizes in the training. Early-stopping is used in our training.
We validate the model every 100-th iteration of the SleepEDF-78 and 500-th in SHHS datasets. We also used the early-stopping when the model didn't update during 1000 periods.

When conducting contrastive learning, it is important to apply augmentation techniques properly.  As augmentations for our contrastive learning, we adopt one of \emph{Amplitude Scaling}, \emph{Amplitude Shift}, \emph{Add Gaussian Noise}, \emph{Band Stop Filter}, \emph{Time Shift} and \emph{Zero-Masking} schemes for raw signals. 
We also employ an augmentation for the spectrogram data but only a simple \emph{Random Noise} augmentation is used for the spectrogram since improperly used augmentations can introduce erroneous modifications that may degrade the quality of features.

\subsection{Experiment result analysis} 

The XSleepNet is composed of a CNN and an RNN network while the proposed model consists of CNN and Transformer model resulting in a slightly larger model size (0.3\% size increase) compared to the XSleepNet. On the other hand, the proposed model achieves a 1.2\% improvement in accuracy compared to the SOTA performance. To further explore the scenario where RNN in XSleepNet is replaced by a Transformer, we modify the XSleepNet model and conduct a performance comparison between the two models. Although the model size increases by 8\%, the modified XSleepNet (XSleepNet+TF) has 0.8\% of accuracy improvement.

To evaluate the training efficiency of our proposed model, we compare the training time of three models: SleepTransformer, XSleepNet, and ours.
Table \ref{table4} shows that the training time of the proposed model is 2.02 (1,005/496) times slower than that of SleepTransformer, whereas XSleepNet demonstrates 2.68 (828/308) times slower than SleepTransformer.
We predict the runtime of our model on a V100 GPU based on the performance ratio observed when the SleepTransformer is executed on two GPUs (it is 496/308 and {\bf 1.61} speedup can be roughly approximated by changing a GPU from RTX3090 to V100). Our model is expected to run on a V100 GPU with an approximate runtime of 624 (1,005/1.61) seconds, which means 1.32 (828/624) times faster training speed than XSleepNet (828 seconds) while showing better accuracy in both the SleepEDF-78 and the SHHS datasets.

\begin{table}[h]
\centering
\caption{Training time (second) per 1000 training steps}
\label{table4}
\begin{tabular}{lccc}
\hline
\textbf{Method} & \textbf{Time} & \textbf{\#Parameters}  & \textbf{GPU} \\
\hline
MC$^2$SleepNet & 1,005 & $5.76 \times 10^6$  & RTX 3090 \\
SleepTransformer & 496 & $3.70 \times 10^6$ & RTX 3090 \\
XSleepNet & 828 & $5.74 \times 10^6$ & Tesla V100 \\
SleepTransformer & 308 & $3.70 \times 10^6$ & Tesla V100  \\
\hline
\end{tabular}
\end{table}

To investigate the detailed performance characteristics of our proposed methods, we conducted an ablation study and evaluated the performance of the different variants of our model. 
In Table \ref{table5}, the `TF only' model refers to a Transformer model, while the `CNN only' model denotes a pure CNN model. These models lack a sequence-level training network and predict a stage class for each epoch solely based on the features extracted from that single epoch, without considering the sequential features between neighboring epochs.
Since they only consider a single epoch for their predictions, their accuracy performances are relatively lower. The `TF+CNN(multi)' model incorporates Transformer and CNN backbones with an additional Transformer-based sequence-level training network, allowing it to extract sequential features from multiple neighboring epochs.
In this model, multi-modal training is utilized with both backbones and sequential features exploited to predict stages, resulting in higher accuracies. When compared to the single-epoch model, performance improvements are observed (83.5\% for SleepEDF-78 and 87.8\% for SHHS 5463).

`TF+CNN+CL+FT' involves applying contrastive learning and fine-tuning. However, employing contrastive learning alone without the masking strategy results in only a 0.2\% performance improvement (83.5\% $\rightarrow$ 83.7\%) compared to that of `TF+CNN(multi)'. Nevertheless, on larger datasets like the SHHS dataset, it demonstrates a 0.6\% (87.8\% $\rightarrow$ 88.4\%) improvement.
`TF+CNN+M+FT' applies a masking strategy without contrastive learning. In this scenario, the performance increases by 0.7\% and 0.6\% on both datasets compared to the `TF+CNN (multi)'.
In the case of `TF+CNN+PT (CL+M)' both contrastive learning and masking are utilized during the pre-training step, where the entire model, including the epoch-level backbones and sequence-level networks, is trained simultaneously, without fine-tuning.
In this model, exploiting the synergy between contrastive learning and masking leads to performance improvements, as shown in Table \ref{table5}. 
Our proposed dual self-supervised learning strategies can achieve, or even surpass, the performance levels obtained by using just a single training method, without the need for a fine-tuning process.

Finally, with the addition of further fine-tuning, the highest accuracy (88.6\%) is observed in `TF+CNN+PT(CL+M)+FT'. It is noteworthy that the separate applications of contrastive learning and masking do not result in performance improvement.

\begin{table}[h]
\setlength{\tabcolsep}{10pt}
\caption{Performance comparison among various model architectures and training schemes. TF: Transformer Backbone, CNN: CNN Backbone, CL: Contrastive Learning in epoch-level, M: Masking prediction in sequence-level with 50\% masking ratio, PT: Pre-training with 50\% masking ratio with `CL' and `M' , FT: Fine-tuning (only sequence-level model part is retrained without masking)}
\label{table5}
\begin{tabular}{l|c|c}
\hline
\textbf{Method} & \textbf{SleepEDF-78} &  \textbf{SHHS 5463} \\
\hline
TF only (single) & 73.8 & 78.8  \\
\hline
CNN only (single) & 79.6 & 84.5  \\
\hline
TF+CNN (multi) & 83.5 & 87.8  \\
\hline
TF+CNN+CL+FT & 83.7 & 88.4  \\
\hline
TF+CNN+M+FT & 84.2 & 88.4  \\
\hline
TF+CNN+PT(CL+M) & 84.4 & 88.3 \\
\hline
TF+CNN+PT+FT & - & 88.6 \\
\hline
\end{tabular}
\end{table}

\begin{figure*}[h]
\centerline{\includegraphics[width = \textwidth]{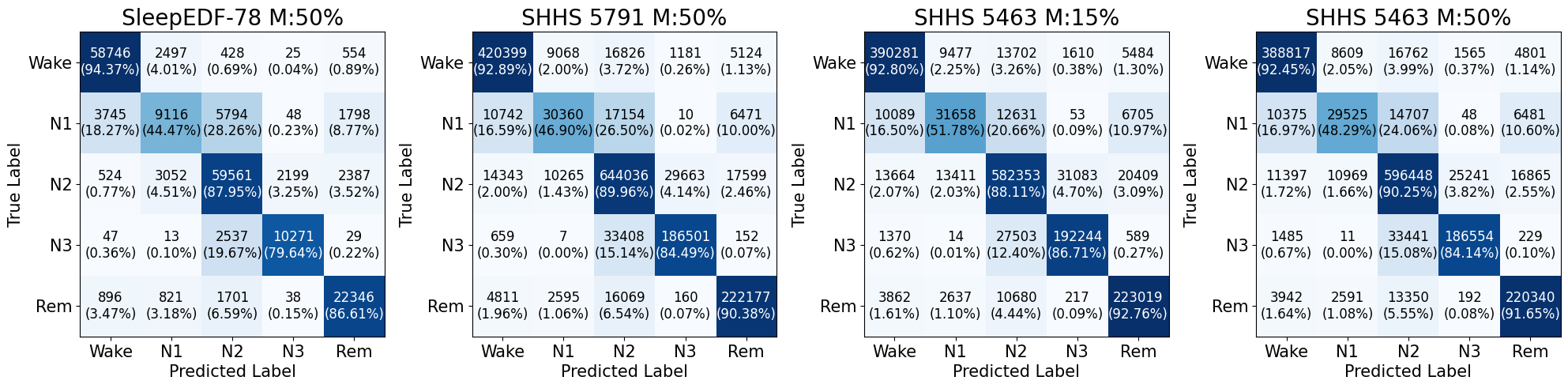}}
\caption{\textbf{Confusion matrices for two datasets and masking ratio (M).}
}
\label{fig3}
\end{figure*}

Controlling the masking ratio is crucial for model accuracy. Increasing the masking ratio to a certain point enhances model accuracy as shown in Table \ref{table6}. Our model shows the highest accuracy (88.495) during the pre-training step when employing a masking ratio (70\%). This suggests that a higher degree of masking plays a crucial role in enhancing the model's ability to learn from incomplete data. 
Overall, the optimal performance (88.593) was observed at the masking ratio of 50\% after fine-tuning. This can be attributed to the fact that after equipping a foundational level of trained knowledge, the model benefits more from a reduced masking ratio. Thus reducing the masking ratio prevents the over-elimination of the extracted feature information and effectively leverages the previously learned patterns. Also, in Fig \ref{fig3}, we see the confusion matrices with the various masking rates.

%


\begin{table}[ht]
\centering
\caption{The performance impacts of a masking ratio and fine-tuning (in the case of an SHHS 5463 dataset).}
\label{table6}
\vspace{-5pt}
\begin{tabular}{ccc}
\hline
\textbf{Masking Ratio} & \textbf{Masking ACC} & \textbf{Fine-tuning} \\
\hline
0\% & - & 88.398 \\
15\% & 88.394 & 88.460 \\
25\% & 88.324 & 88.453 \\
30\% & 88.451 & 88.513 \\
40\% & 88.261 & 88.476 \\
50\% & 88.324 & \textbf{88.593} \\
60\% & 88.292 & 88.426 \\
70\% & \textbf{88.495} & 88.506 \\
75\% & 88.398 & 88.495 \\
85\% & 88.354 & 88.515  \\
\hline
\end{tabular}
\end{table}

 

In Table \ref{table2}, we compare our model with various models to demonstrate the efficacy of our approach. Our model outperforms all the previous works in all sleep stages except for the `Wake' stage in both datasets.  
The ambiguity and similarity between the NREM1 and NREM2 classes pose challenges in sleep staging. It is even challenging to classify them accurately in our model.
In general, the NREM2 class occupies a larger proportion of data samples compared to the NREM1 class and it makes the prediction of the NREM1 class highly challenging.
Our method outperforms other similar models such as XSleepNet by 0.4\%. More precisely, we exceed performance by 0.7\%, 0.5\%, 1.9\%, and 2.7\% in the NREM1, NREM2, NREM3, and REM classes, respectively, on the SleepEDF-78 Dataset.
The `Input' sub-column of the `Method' column of Table \ref{table2} shows what types of data are used. `RS' and 'SP' mean raw signal and spectrogram data, respectively.

\begin{table*}
\centering
\caption{\textbf{Performance comparison of sleep stage classification models on SleepEDF-78 and SHHS Datasets.}
}
\label{table2}
\renewcommand{\arraystretch}{1.5}
\begin{tabular}{cccc|ccc|ccccc}
\hline\hline
\multicolumn{4}{c|}{\textbf{Method}} & \multicolumn{3}{|c|}{\textbf{Overall Metrics}} & \multicolumn{5}{|c}{\textbf{Per-class F1 score}} \\
\hline
Dataset & System & Input & Subjects & ACC & MF1 & k & W & N1 & N2 & N3 & REM \\
\hline

SHHS & MC$^2$$\text{SleepNet}^{50\%}$ (Ours) & RS + SP & 5463 & \textbf{88.59} & \underline{82.1} & \textbf{0.841} & \underline{93.0} & \underline{52.3} & \textcolor{red}{\textbf{89.3}} & \underline{85.7} & \textcolor{red}{\textbf{90.1}} \\

SHHS & MC$^2$$\text{SleepNet}^{15\%}$ (Ours) & RS + SP & 5463 & \underline{88.46} & \textbf{82.3} & \underline{0.84} & 92.9 & \textcolor{red}{\textbf{53.5}} & \underline{89.1} & \textcolor{red}{\textbf{86.0}} & \underline{89.8} \\

SHHS & L-SeqSleepNet \cite{phan2023seqsleepnet} & SP & 5463 & 88.35 & 81.4 & 0.838 & \textbf{93.1} & 51.1 & 89.0 & 84.9 & \underline{89.8} \\

\hline

SHHS & MC$^2$$\text{SleepNet}^{50\%}$ (Ours) & RS + SP & 5791 & \textbf{88.5} & \textbf{81.7} & \textbf{0.837} & \textbf{93.1} & \textbf{51.9} & \textbf{89.2} & \underline{85.1} & \textbf{89.3} \\

SHHS & CoRe-Sleep \cite{kontras2023coresleep} & SP & 5791 & \underline{88.2} & \underline{81.0} & \underline{0.834} & - & - & - & - & - \\

SHHS & SleePyco \cite{lee2024sleepyco} & RS & 5793 & 87.9 & 80.7 & 0.830 & \underline{92.6} & 49.2 & \underline{88.5} & 84.5 & \underline{88.6} \\

SHHS & SleepTransformer \cite{phan2022sleeptransformer} & SP & 5791 & 87.7 & 80.1 & 0.828 & 92.2 & 46.1 & 88.3 & \textbf{85.2} & \underline{88.6} \\

SHHS & L-SeqSleepNet \cite{phan2023seqsleepnet} & SP & 5791 & 87.6 & 80.3 & 0.825 & 92.4 & 48.6 & 88.2 & 83.9 & 88.5 \\

SHHS & XSleepNet \cite{phan2021xsleepnet} & RS + SP & 5791 & 87.6 & 80.7 & 0.826 & 92.0 & \underline{49.9} & 88.3 & 85.0 & 88.2 \\

SHHS & SeqSleepNet \cite{phan2019seqsleepnet} & SP & 5791 & 86.5 & 78.5 & 0.81 & 91.8 & 49.1 & 88.2 & 83.5 & 88.2 \\

\hline

SleepEDF-78 & MC$^2$$\text{SleepNet}^+$ (Ours) & RS + SP & 78 & \textbf{84.6} & \textbf{79.1} & \textbf{0.787} & 93.1 & \textbf{50.6} & \textbf{86.5} & \textbf{80.6} & \underline{84.5} \\

SleepEDF-78 & SleePyco \cite{lee2024sleepyco} & RS & 78 & \textbf{84.6} & \underline{79.0} & \textbf{0.787} & \textbf{93.5} & \underline{50.4} & \textbf{86.5} & \underline{80.5} & 84.2 \\

SleepEDF-78 & MC$^2$$\text{SleepNet}^{50\%\Gamma}$(Ours) & RS + SP & 78 & \underline{84.4} & 78.5 & \underline{0.784} & \underline{93.3} & 49.3 & \textcolor{red}{\textbf{86.5}} & 78.8 & \textcolor{red}{\textbf{84.6}} \\

SleepEDF-78 & MC$^2$$\text{SleepNet}^{15\%\Gamma}$(Ours) & RS + SP & 78 & \underline{84.4} & 78.7 & \underline{0.784} & \underline{93.3} & \textcolor{red}{49.4} & \underline{86.4} & \textcolor{red}{80.0} & 84.2 \\

SleepEDF-78 & XSleepNet \cite{phan2021xsleepnet} & RS + SP & 78 & 84.0 & 77.9 & 0.778 & \underline{93.3} & 49.9 & 86.0 & 78.7 & 81.8 \\

SleepEDF-78 & SeqSleepNet \cite{phan2019seqsleepnet} & SP & 78 & 82.6 & 76.4 & 0.760 & - & - & - & - & - \\

SleepEDF-78 & SleepTransformer \cite{phan2022sleeptransformer} & SP & 78 & 81.4 & 74.3 & 0.743 & 91.7 & 40.4 & 84.3 & 77.9 & 77.2 \\

SleepEDF-78 & SleepEEGNet \cite{EEGSleepNet} & RS & 78 & 80.0 & 73.6 & 0.730 & 91.7 & 44.1 & 82.5 & 73.5 & 76.1 \\

\hline\hline
\end{tabular}
\end{table*}

The proposed MC$^2$SleepNet demonstrates improved performance on a larger dataset as well. While L-SeqSleepNet \cite{phan2023seqsleepnet} achieves SOTA performance using a relatively longer sequence length (L=200), Our MC$^2$SleepNet surpasses it while employing a relatively shorter sequence length.
Our MC$^2$SleepNet achieves the highest ACC (88.59\%) and MFI (82.3\%) scores when compared to the previous works as presented in Table \ref{table2}.
It is noteworthy that the F1 score for the minor classes, NREM1 (3.8\% of the SHHS dataset) and NREM3 (6.7\% of the SleepEDF-78 dataset), are enhanced when a 15\% masking ratio is used. Conversely, the major classes, NREM2 and REM,  achieve best results with a 50\% masking ratio, consistently as indicated by red color in Table \ref{table2}.
%
%
By adjusting the masking ratio readily, the proposed model can be tailored partly to prioritize some classes

In Table \ref{table2}, the `+' sign indicates that the models are pre-trained on the SHHS dataset with 5463 patients and a fine-tuning is conducted with SleepEDF-78. On the other hand, to prevent the over-fitting problem in the small size dataset, we conduct only the pre-training step symboled ${\Gamma}$ in the table.

\section{Conclusion} 
In this paper, we introduced an MC$^2$SleepNet for sleep stage classification which aimed to achieve effective collaborative learning of the features extracted from multi-modal data through CNN and Transformer architectures.
By utilizing multi-modal data samples, and putting the data of each modal into different networks, our model overcame the limited inspection and exploration on the feature representation space that could be caused by limited views.
A contrastive learning technique is employed to leverage the features extracted through CNN and Transformer backbones from these multi-modal data sources.
Additionally, we have developed a 'Cross-Masking' scheme based on a cross-attention mechanism for sequence-level training, which enhances performance in classifying both minor and major classes.
%
Our MC$^2$SleepNet has achieved the \textit{state-of-the-art} performance with an accuracy of both 84.6\% on the SleepEDF-78 and 88.6\% accuracy on the SHHS dataset. This demonstrates that our proposed network is generalized effectively across small and large datasets.


\bibliographystyle{ACM-Reference-Format}
\bibliography{draft_sample}

\end{document}